\documentclass[%
 reprint,
 amsmath,amssymb,
 aps,
 superscriptaddress,
]{revtex4-2}

\usepackage{eqnarray}
\usepackage{xcolor}
\usepackage{graphicx}
\usepackage{dcolumn}
\usepackage{bm}
\usepackage{hyperref}

\begin{document}
\title{Ultra-coherent nanomechanical resonators based on density phononic crystal engineering}

\author{Dennis H\o j}
	\email{denho@fysik.dtu.dk}
	\affiliation{Center for Macroscopic Quantum States (bigQ), Department of Physics, Technical University of Denmark, Fysikvej, 2800 Kgs. Lyngby, Denmark}

\author{Ulrich Busk Hoff}
\email{ulrich.Hoff@fysik.dtu.dk}
	\affiliation{Center for Macroscopic Quantum States (bigQ), Department of Physics, Technical University of Denmark, Fysikvej, 2800 Kgs. Lyngby, Denmark}

\author{Ulrik Lund Andersen}
	\email{ulrik.andersen@fysik.dtu.dk}
	\affiliation{Center for Macroscopic Quantum States (bigQ), Department of Physics, Technical University of Denmark, Fysikvej, 2800 Kgs. Lyngby, Denmark}




\begin{abstract}
 Micro- and nanomechanical systems with exceptionally low dissipation rates are enabling the next-generation technologies of ultra-sensitive detectors and quantum information systems. New techniques and methods for lowering the dissipation rate have in recent years been discovered and allowed for the engineering of mechanical oscillators with phononic modes that are extremely well isolated from the environment and thus possessing quality factors close to and beyond 1 billion. A powerful strategy for isolating and controlling a single phononic mode is based on phononic crystal engineering. Here we propose a new method for phononic crystal engineering of nanomechanical oscillators that is based on a periodic variation of the material density. To circumvent the introduction of additional bending losses resulting from the variation of material density, the added mass constitutes an array of nanopillars in which the losses will be diluted. Using this novel technique for phononic crystal engineering, we design and fabricate corrugated mechanical oscillators with quality factors approaching one billion in a room temperature environment. The flexibility space of these new phononic crystals is large and further advancement can be attained through optimized phononic crystal patterning and strain engineering via topology optimization. This will allow for the engineering of mechanical membranes with quality factors approaching 10 billion. Such extremely low mechanical dissipation rates will enable the development of radically new technologies such as quantum-limited atomic force microscopy at room-temperature, ultra-sensitive detectors of dark matter, spontaneous waveform collapses, gravity, and high-efficiency quantum information transducers.  
\end{abstract}

\maketitle

Phononic crystal engineering is an extremely powerful method for controlling the phononic properties of material systems in much the same way as photonic crystals are able to control light~\cite{Maldovan2013,Vasileiadis2021}. The method is being used to engineer material systems for controlling heat and sound both at the macro- and the micro-scale, for example with applications in the design of loudspeakers and microphones to control the direction of sound~\cite{Song2020}, in buildings and vehicles to isolate sound, and in communication technologies to enable signal processing at 5G frequencies. In particular, in recent years, phononic crystals have been used to control and isolate specific phononic modes of nano- and micromechanical resonators, protecting them from the surrounding thermal heat bath~\cite{Yu2014,Tsaturyan2014,Ghadimi2017b,Tsaturyan2017,Ghadimi2017,Beccari2021}. They greatly outperform other resonator designs~\cite{Verbridge2006,Reinhardt2016,Norte2018,Pratt2021} and have given rise to a completely new generation of ultra-coherent mechanical oscillators with quality factors reaching one billion at room temperature.

The effect of enhancing the quality factor of mechanical resonators using phononic crystals is two-fold. First, as a result of the periodic phononic structure, an acoustic bandgap is formed in which phononic modes are trapped and thus unable to radiate into the surrounding environment that otherwise would cause dissipation~\cite{Yu2014,Tsaturyan2014,DeJong2022}. Second, the phononic crystal structure can be also engineered to reduce the phononic mode curvature of the resonator at the clamping points to the substrate \cite{Schmid2011,Villanueva2014}, which will lead to a decrease in the intrinsic dissipation rate for highly stressed systems – a technique that is known as soft-clamping~\cite{Tsaturyan2017,Ghadimi2017,Beccari2021,Bereyhi2022,Shin2022,Bereyhi2022b}. Phononic crystal membranes and strings exploiting these effects in harmony have been fabricated, and the membranes have been used in a series of optomechanical experiments~\cite{Rossi2018,Mason2019,Chen2020}.       

A phononic crystal in a mechanical membrane is traditionally engineered by mapping a periodic pattern of holes into a pre-stressed silicon membrane. This produces a periodic variation in the stress of the material across the membrane, which in turn creates a phononic bandgap with isolated modes that are associated with the vibrations at a central "defect" of the pattern. As mentioned above, this leads to acoustic isolation as well as soft-clamping of a particular mode, and thus very large quality factors. However, the stress of the material at the location of the high-quality mode is low which means that the quality-enhancing effect of dissipation dilution\cite{Gonzalez1994,Verbridge2006,Federov2019b} is not optimally used, and therefore eventually bounds the quality factor of the system. The technique of strain engineering can in some geometries be applied to co-locate a high stress region with the mechanical motion~\cite{Ghadimi2017} but in this particular phononic membrane, it seems not possible. 

\begin{figure*}[t]
\centering
\includegraphics{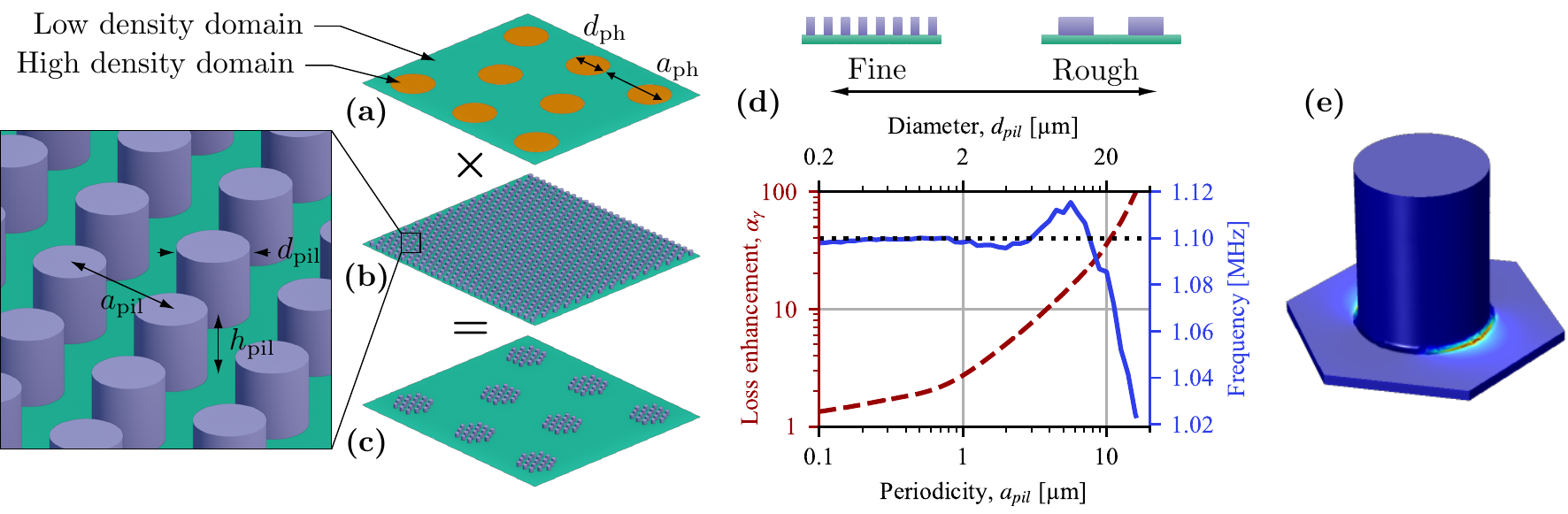}
\caption{\textbf{Concept illustration:} (a) The desired phononic crystal pattern of high and low material density regions. To approximate this behaviour the distribution is used as a mask on-top an even distribution of nano-pillars shown in (b). The end result is an engineered distribution of nano-pillars based on any desired material distribution as illustrated in (c). The zoomed-out inset of (b) highlights the relevant pillar parameters. (d) The increased losses relative to a hypothetical membrane (without pillars) of increased material density while keeping the total mass constant. It assumes a 20 nm thick stoichiometric silicon nitride membrane vibrating at 1.1 MHz. The results are practically identical to the other studied frequencies. (e) Illustration of evanescent bending losses at the pillar base, estimated using 3D simulations.}
\label{fig:concept_illustration}
\end{figure*}

In this Letter, we propose a new type of phononic crystal engineering of a nano- and micromechanical resonator that offers a larger flexibility in the design of periodic patterns and in which the stress can be engineered. Our strategy for phononic crystal nano-engineering is based on the creation of a periodic variation in the mass density across the membrane. More specifically, we propose and demonstrate the mapping of a periodic hexagonal pattern with a central defect onto a membrane that consists of an array of nanopillars, resulting in a membrane with periodic pattern of cell domains each containing an array of nanopillars as illustrated in Fig.\ref{fig:concept_illustration}a-c (note that the central defect is not shown in the figure). 
Such a density phononic membrane enables the formation of a frequency bandgap with modes that are highly shielded from the substrate (acoustic filtering) and it reduces the curvature of the mode amplitude towards the clamping points, thereby lowering the intrinsic dissipation losses (soft-clamping). Moreover, the use of a fine mesh of nanopillars for controlling the mass density ensures low intrinsic bending dissipation at the pillars. We therefore expect the density phononic membranes to exhibit ultra-low dissipation rates.

Intuitively, the simplest strategy for phononic crystal engineering via material density variation is to periodically add mass by depositing large pillars representing the cells of the phononic crystal pattern. Such a structure will in principle create a bandgap (similar to a holey structure) but the structure will be too stiff and it will be strongly affected by boundary bending losses at each of the domain walls. However, this stiffness and dissipation effect can be largely reduced by patterning a fine mesh of nanopillars within each of the cells. With such a strategy, the density contrast required for phononic engineering is kept while the bending loss is significantly reduced. This reduction is caused by the dilution of the loss into many nanopillars each of which produces almost negligible loss due to their small diameter in comparison to the wavelength of the phononic mode. We simulate the potential loss enhancement (see methods) by considering a hexagonal pattern of pillars on a membrane with the effective density
\begin{equation}
    \rho_\textrm{eff}=\rho\left(1+\frac{\pi\sqrt{3}h_\textrm{pil}d_\textrm{pil}^2}{6t a_\textrm{pil}^2}\right)
\end{equation}
where $\rho$ is the density of the material (assuming the same material for membrane and pillars), while $a_\textrm{pil}$ and $h_\textrm{pil}$ are the diameter and height of the pillars, $d_\textrm{pil}$ is the distance between them (see Fig.~\ref{fig:concept_illustration}b) and $t$ is the thickness of the membrane. The results of the simulations are presented in Fig.~\ref{fig:concept_illustration}d where we plot the loss enhancement $\alpha_\gamma$ of a membrane patterned with a hexagonal array of pillars against the periodicity, $a_\textrm{pil}$, or pillar size, $d_\textrm{pil}$ (assuming $a_\textrm{pil}=2d_\textrm{pil}$). It is clear that the intrinsic loss of the membrane is dramatically reduced by the introduction of a nanopillar array, and for very small pillars, the loss contribution is effectively negligible. 

Since the phononic crystal pattern is created by material density variation, there exists a large flexibility in defining its form and shape. Two examples are a binary and a sinusoidal pattern where the material density is either changed abruptly or smoothly across the membrane following a certain periodicity as shown in Fig.~\ref{fig:phononic_analysis}c. 
Analytical expressions for these two patterns can be found in the supplementary material, and in the following we interrogate the binary pattern while the results of the sinusoidal pattern are discussed in the supplementary material. We assume the cell periodicity to be given by $a_\textrm{ph}$, the diameter of each cell to be $d_\textrm{ph}$ and the ratio $\alpha_w=d_\textrm{ph}/a_\textrm{ph}$. Furthermore we define the density contrast parameter, $g=\rho_\textrm{eff}/\rho$.

The optimization of the density phononic crystal is a trade-off between the size of the bandgap, which should be as large as possible to encapsulate the mode of interest, and the dissipation of an isolated mode, which should be as low as possible. In Fig.~\ref{fig:phononic_analysis}a-b we present the results of simulating the relative bandgap width and the quality factor of the isolated mode assuming that intrinsic losses are dominating. It is clear that the bandgap becomes wider as the density contrast, $g$, increases as expected, and that an optimum is obtained for a specific relative distribution width, $\alpha_w$, of around 0.4 (corresponding to $a_\textrm{ph}=2.5d_\textrm{ph}$). However, contrarily, the quality factor for the isolated mode is decreasing as the density contrast is increased. This is caused by the fact that a larger material density will lead to a smaller phase velocity and therefore a shorter wavelength of the propagating phononic mode. A shorter wavelength relative to the fixed size of the nanopillars will naturally lead to larger bending losses at the nanopillars, and thus a reduced quality factor. In the following, we choose a contrast of $g=5$ which is a good compromise between having bandgap of reasonable size and a large quality factor. In addition, based on these simulations, in the following we choose $\alpha_w=0.4$. 

To trap an isolated mode in the membrane, a defect is introduced simply by removing one of the cell domains. Such a disturbance in the phononic crystal allows some modes to occur in the bandgap which will be spatially well confined at the defect. However, the resonance frequency of the high quality mode is not centered in the bandgap but at the lower end of it which might lead to some mode leakage. To avoid this effect, we increase the frequency of the mode by lowering the material density of the phononic cells located just next to the defect. By choosing the contrast of these inner cells to $g=3.9$ or the relative size to $\alpha_w=0.32$, the frequency will be centered in the bandgap. In Fig.~\ref{fig:phononic_analysis}, we show the contour of the density contrast (Fig.~\ref{fig:phononic_analysis}d), the displacement amplitude of the high quality mode (Fig.~\ref{fig:phononic_analysis}e) and the dispersion relation that clearly exhibits the bandgap (Fig. \ref{fig:phononic_analysis}f). Interestingly, we note that it generates a full bandgap in contrast to stress-induced phononic membranes which produces a pseudo-bandgap with in-plane modes~\cite{Tsaturyan2017}. Finally, in Fig.~\ref{fig:phononic_analysis}c we present the results of the simulated quality factor (based on intrinsic losses) for different number of layers of the phononic crystal cells as well as three different values for the periodicity of the phononic pattern ($a_\textrm{ph}={199, 221, 253}\ \mu m$). In these simulations we assume no added losses from the pillars ($\alpha_\gamma = 1$) .  
The effect of soft-clamping is clear. As more layers are added, the mode field curvatures at the boundary is gradually reduced which in turn leads to lower intrinsic dissipation rates, thereby increasing the quality factor. We see that at least 7 layers of phononic cells are required for saturating the quality factor to a value above $10^9$. 

Our next step is to experimentally validate the expected quality of our proposed density phononic crystal. Towards this end, we fabricate membrane structures for different pillar periodicity and for different numbers of phononic cell layers. We fabricate the membranes from two thin layers of high-stress (about 1GPa) silicon nitride films grown on a silicon wafer using low-pressure chemical vapor deposition (CVD) and plasma-enhanced CVD, respectively. Using UV lithography and reactive ion etching the nanopillar pattern was transferred to the front-side while a window for releasing the resonator was transferred to the back-side. More details on the fabrication process can be found in the method section.      

We characterize the fabricated membranes by determining their frequency power spectrum as well as their quality factors using an interferometric ringdown measurement setup operating in ultrahigh vacuum to avoid air damping (see methods). For the ringdown measurements, we excite the membrane by the exertion of radiation pressure force from a light beam modulated at the frequency of the mode under investigation, turn off the modulation and subsequently record the decay of the membrane’s motion using the same laser mode. An example of the measured power spectrum is illustrated in Fig.~\ref{fig:experimental_results}c and shows the clear existence of a bandgap around the confined mode at 1.4 MHz. The result of the ringdown measurement of this particular mode is presented in Fig.~\ref{fig:experimental_results}d and yields a quality factor of 5.65$\times$10$^8$.  

In Fig.~\ref{fig:experimental_results}e we present the results for the measured quality factors of the bandgap mode with different number of layers of phononic cells. As the number of cell layers increases, the quality factor improves since the increasing number of layers will gradually strengthen the effects of soft-clamping and acoustic shielding against phonon tunneling loss. When the acoustic shielding is weak (that is, when only a few cell layers are present), the coupling to the surrounding substrate mode becomes stochastic resulting in a large spread of the quality factors. For a large number of cell layers ($N_{ph}\geq 9$), the acoustic shielding effect becomes more effective and the spread is low. In Fig.~\ref{fig:experimental_results}e, we compare the measured data with theory that accounts for intrinsic losses but neglecting the phonon tunneling losses, and we clearly observe reasonable agreements to the best measured values.  

Results for the quality factor of the bandgap mode for different nanopillar periodicity, $a_{pil}$, are presented in Fig.~\ref{fig:experimental_results}f. As we kept the phononic crystal pattern constant, a change in the periodicity effectively leads to a change in the nanopillar diameter ($d_{pil}=a_{pil}/2$) and density. As expected, the quality factor increases as we lower the periodicity (and thus the size of the nanopillars) as this will dilute the effect of bending losses at the pillars. We expect the quality factors to be limited by the intrinsic bending losses since we have chosen a large number of cell layers, $N_{ph}=8$, diminishing the phonon tunneling losses. Indeed, by using a model that accounts only for intrinsic losses we obtain good agreement with the measured data as shown in Fig.~\ref{fig:experimental_results}f. 

The smallest pillars that we were able to fabricate with the present fabrication method was around one micron which yields the maximum quality factor of $9.1\times 10^8$, and a Qf-product of $1.3\times10^{15}$ (corresponding to about 216 quantum coherent oscillations) which we believe is the largest reported Qf-product at room temperature for membranes and similar to the best measured perimeter mode of a polygon-shaped resonator~\cite{Bereyhi2022b}. Moreover, according to our model, the quality factor will increase to beyond one billion by further decreasing the periodicity. Furthermore, due to the structural functionality of our system, it is possible to enhance the strain of the membrane and thereby further decreasing the intrinsic losses through dissipation dilution. One particular strategy for increasing the stress is to use the design depicted in Fig.~\ref{fig:twodim_strain_engineering} where the stress of the central pad (on which the density phononic crystal could be located) is reaching approximately 3 GPa. In addition to improving the quality factor further by strain engineering, it is also possible to improve it by choosing optimised variations of the material density. E.g. in the supplementary material we have considered a sinusoidal variation of the material density by which the quality factor is slightly improved, and we expect even further improvements by optimizing the density variation using for example the mathematical framework of topology optimization~\cite{Hoej2021}.     

In conclusion, we proposed and demonstrated a new method for phononic crystal engineering of micro- and nanomechanical membranes that realizes acoustic shielding and soft-clamping, and that leaves a lot of flexibility for parameter optimization and strain-engineering. The fabricated devices exhibits quality factors as high a $9.1\times 10^8$ at room temperature, and with further optimization of pillar size (see Fig.~\ref{fig:phononic_analysis}c), pattern design (see Fig. 8b in supplementary materials) and strain-engineering (see Fig.~\ref{fig:twodim_strain_engineering}b), we anticipate quality factors approaching 10 billion at room temperature. The method is material independent and we expect it can be combined with crystalline materials for even greater quality factors~\cite{Romero2020,Beccari2021}. With such quality factors, it is possible to bring the membrane into the quantum ground state at room temperature via feedback cooling (possibly without the use of a cavity) and to eventually prepare non-classical mechanical states such as squeezed and entangled states with applications in quantum repeaters~\cite{Fiaschi2021} and sensors. Even more intriguingly, the high-quality of our membranes might eventually facilitate the preparation of quantum non-Gaussian mechanical states through the interaction with two-level systems such as defect centers in diamond~\cite{Lee2017} or superconducting qubits~\cite{Clerk2020}, and the realization of quantum transduction to coherently connect two disparate systems with close to unity efficiency~\cite{Barzanjeh2022,Lauk2020,Delaney2022}. Finally, in the current work, we have focused on the improvement of the quality factor, and the Qf-product, which is of high relevance for many of the applications mentioned above. However, in a future work, it would be interesting to optimize the Q/f-fraction for applications in atomic force microscopy or optimize the $Q\times m$-product (where m is the effective mass) as required for testing alternative models of quantum mechanics~\cite{Manley2021,Nimmrichter2014,Westphal2021}.


\begin{figure*}[t]
\centering
\includegraphics{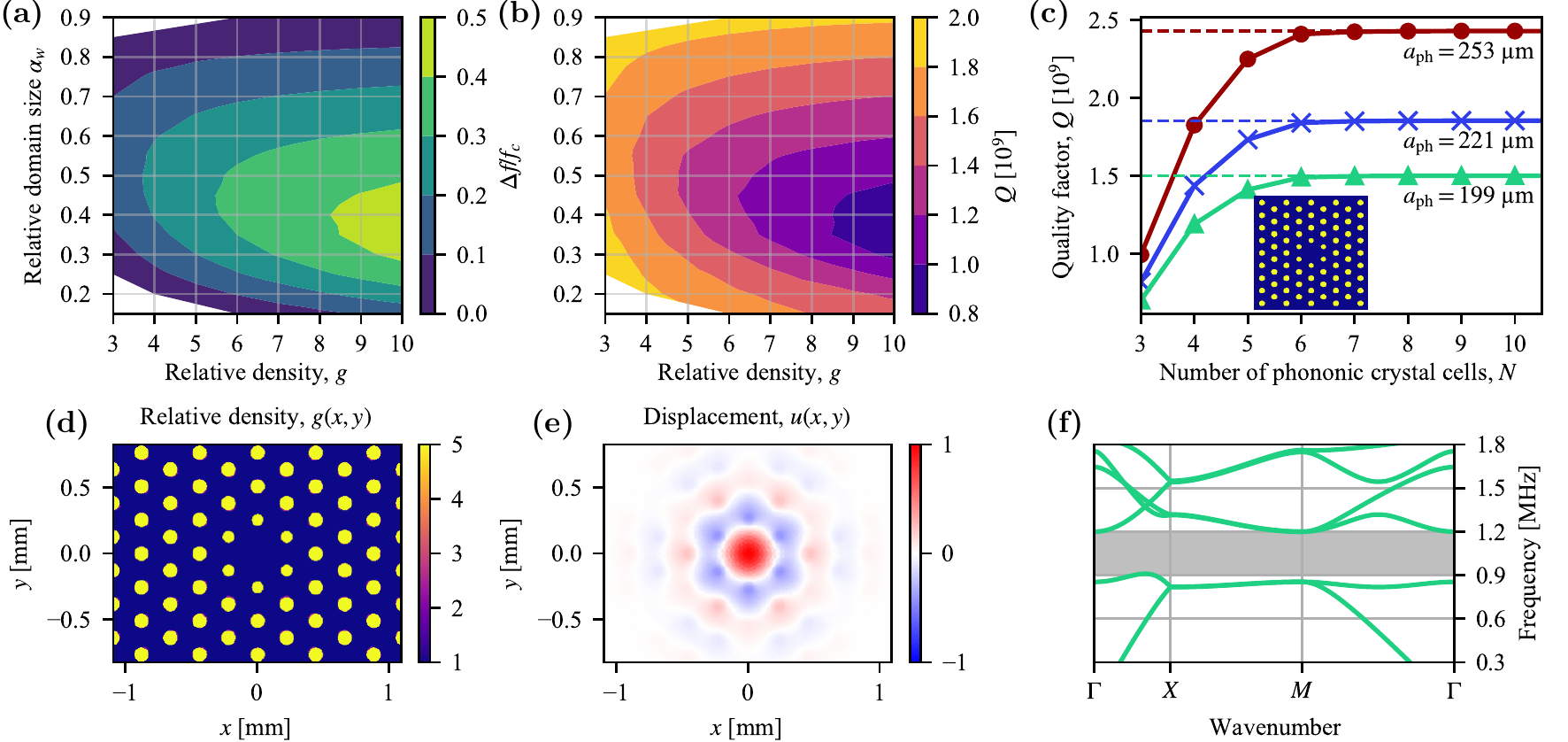}
\caption{\textbf{Phononic analysis:} (a) and (b) show the relative bandgap, $\Delta f/f_c$, and quality factor, $Q$, respectively, versus the phononic crystal contrast, $g$, and the relative domain size, $\alpha_w$ (fixing the periodicity to $a_{ph} = 253 \ \textrm{\textmu{}m}$ membrane. (c) Predicted quality factors versus number of isolating phononic crystal cell layers. The insert illustrates a membrane with $N=4$. In (b) and (c), the Q factor was numerically estimated without accounting for the additional pillar loss (assuming very small pillars) and without accounting for acoustic radiation loss. (d) and (e) illustrate the density distribution and the modeshape of the phononic membrane, respectively. (f) The band structure of the phononic crystal, where the shaded area highlights the bandgap.}
\label{fig:phononic_analysis}
\end{figure*}

\begin{figure*}[t]
\centering
\includegraphics{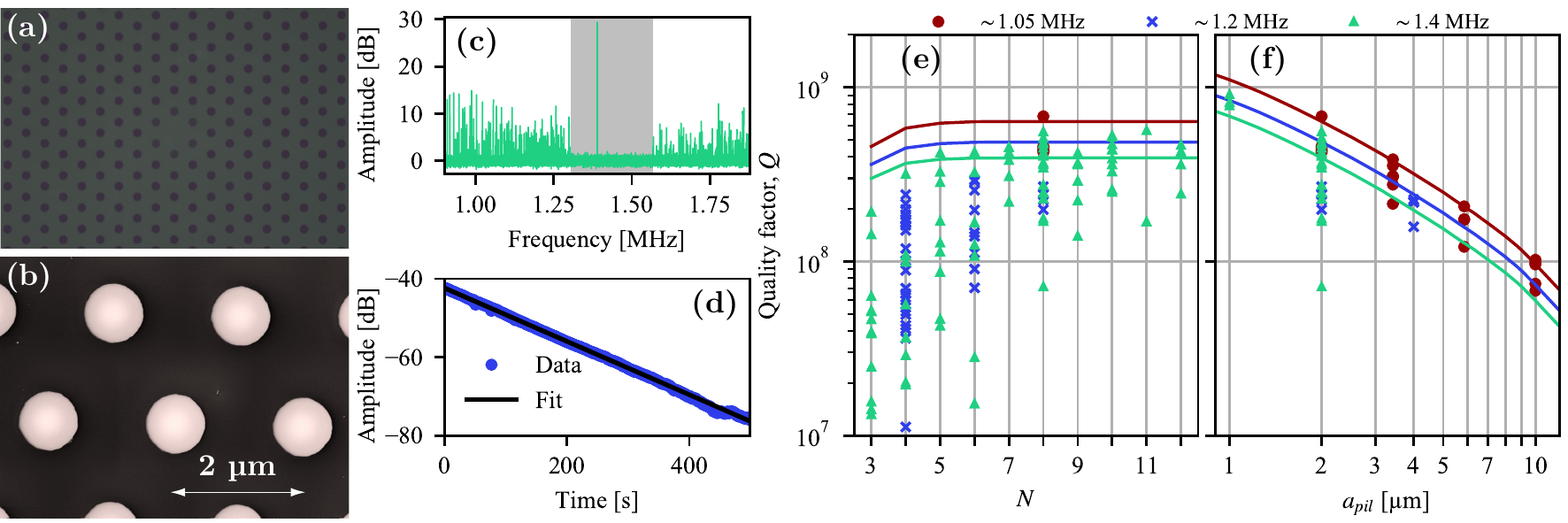}
\caption{\textbf{Experimental results:} (a) Microscope image of the fabricated density phononic membrane. (b) SEM image of the nanopillars of a specific membrane. (c) Spectrum of a density phononic membrane where the band-gap region is highlighted by a shaded grey. (d) An example of a ringdown measurement of a specific membrane exhibiting a Q factor of 5.65$\times$10$^8$. (e) Measured quality factors versus number of phononic cells $N$ for $a_\textrm{pil} = 2 \ \textrm{\textmu m}$. (f) Measured quality factor versus pillar periodicity $a_\textrm{pil}$ for $N=8$. The solid lines correspond to fitted simulation results based on the best samples.}
\label{fig:experimental_results}
\end{figure*}

\begin{figure}[t]
\centering
\includegraphics{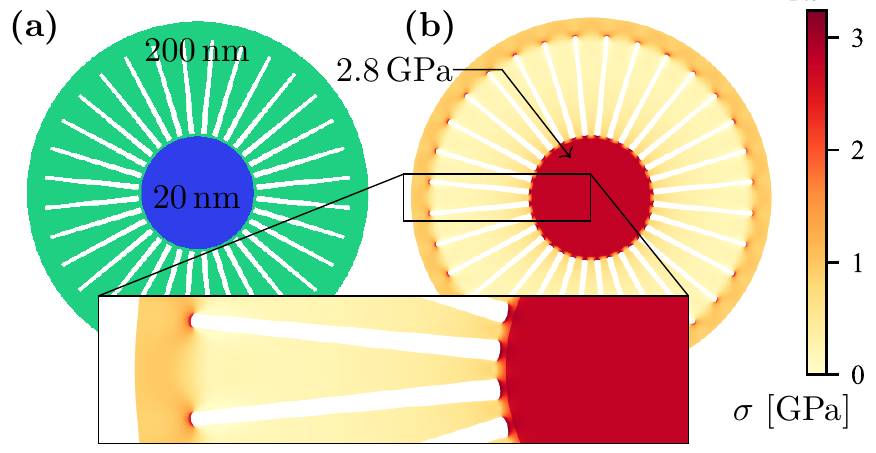}
\caption{\textbf{2D strain engineering:} (a) Theoretical design of a stoichiometric silicon nitride membrane consisting of two different thicknesses (20nm and 200nm) which enables stress enhancement of two-dimensional membranes. (b) Illustration of the resulting von Mises tensile stress distribution.}
\label{fig:twodim_strain_engineering}
\end{figure}

\section{Methods}
{\it Simulations:} The simulations of the membranes and nanopillars were performed in COMSOL Multiphysics. The nanopillars were simulated in full 3D for a single pillar using Floquet periodic boundary conditions. The wavenumber was defines as $k = 2\pi f\sqrt{g\rho/\sigma}$ (where $\sigma$ is the tensile stress and $f$ is the frequency) and the damping was extracted in post-processing by integrating the strains over the entire volume. The loss factor and other material parameters were assumed constant and isotropic. Based on these results, the membranes were modelled in 2D assuming a uniform tensile stress distribution and by adjusting the material density using $\rho \to \rho(x,y) = g(x,y)\rho$. The losses in the density-increased domains are scaled based on the chosen pillar periodicity $a_\textrm{pil}$ and pillar simulation results. The quality factor $Q = 2\pi W/\Delta W$ is then estimated in post-processing using the kinetic energy \begin{equation}
W = 2\pi^2f^2 t\iint \rho(x,y) u^2(x,y) \ \mathrm{d}x\mathrm{d}y
\end{equation}
and dissipation via bending 
\begin{multline}
    \Delta W = \frac{\pi Et^3}{12(1-\nu^2)} \iint \phi(x,y) 
    \\
    \cdot
    \left(\frac{\partial^2u(x,y)}{\partial x^2} + \frac{\partial^2u(x,y)}{\partial y^2)}\right)^2 \ \mathrm{d}x\mathrm{d}y    
\end{multline}
where $t, E, \nu$ are the thickness, Young's modulus and Poisson's ratio of the resonator material. $u(x,y)$ is the mode shape and $\phi(x,y)$ is the effective loss angle of the membrane which is affected by the choice of material, its thickness, and the presence of pillars.

{\it Fabrication:} Double-sided polished Silicon wafers (of 500 \textmu{}m thickness) were initially deposited with 20 nm stoichiometric silicon nitride via low-pressure chemical vapor deposition, followed by a layer of 1000 nm silicon nitride using plasma-enhanced chemical vapor deposited (PECVD). The pillar pattern was transferred onto a spincoated layer of photo-resist via UV-lithography followed by reactive ion ethcing almost through the PECVD layer. Another round of lithography and reactive ion etch was performed on the backside to define the membrane window. Finally, a short etch in buffered hydrofluoric acid followed by an etch in potassium hydroxide ensured the removal of the residual PECVD layer and release of the membrane. The membranes were then cleaned in hydrochloric acid followed by sulfiric acid mixed with ammoniumpersulfate and later dried in ethanol fumes.

{\it Optical characterization:} The phononic spectra as well as the Q factors are measured using optical interferometry at 1550 nm. The membrane under study is placed inside a vacuum chamber with pressure $<$10\textsuperscript{-7}mBar and excited with a modulated laser beam reflecting off its central part. Once excited, the laser modulation is turned off, and the phase shift of the reflected beam is measured as a function of time with a phase-locked homodyne detector. The signal is used to record the phononic mode spectrum as well as the amplitude decay of the phase modulation with specific examples shown in Fig.~\ref{fig:experimental_results}c and Fig.~\ref{fig:experimental_results}d.

\bibliography{Ref}

\end{document}